\documentclass[preprint,
 amsmath,amssymb,
 prd,
floatfix,
]{revtex4-2}

\usepackage{graphicx}
\usepackage{dcolumn}
\usepackage{textcomp}
\usepackage{gensymb}
\usepackage{bm}
\usepackage{subfigure}
\usepackage{subfloat}
\usepackage{tabularx}
\usepackage[toc,title,page]{appendix}
\usepackage{hyperref}

\providecommand{\sorthelp}[1]{}
\newcommand{\unit}[1]{\ensuremath{\, \mathrm{#1}}}

\newcommand{\dd}{\mathrm{d}}
\newcommand{\Hi}{\mathcal{H}_\mathrm{i}}
\newcommand{\Ha}{\mathcal{H}_\mathrm{a}}

\begin{document}

\title{Directional Variations of Cosmological Parameters from the \textit{Planck} CMB Data}

\author{S. Yeung}
\email{terryys@link.cuhk.edu.hk}
 
\author{M.-C. Chu}
\email{mcchu@phy.cuhk.edu.hk}
\affiliation{Department of Physics and Institute of Theoretical Physics, The Chinese University of Hong Kong\\Shatin, Hong Kong}

\date{\today}

\begin{abstract}
Recent observations suggest that there are violations of the isotropy of the universe at large scales, an important part of the cosmological principle. In this paper, we use the Cosmic Microwave Background (CMB) data to search for spatial variations of the cosmological parameters in the $\Lambda\mathrm{CDM}$ model. 
We fit the \textit{Planck} temperature angular power spectrum $\mathcal{C}^{TT}_\ell$ for 48 different half-skies, centering on  48 different directions, to search for directional dependences of the standard cosmological parameters. 
There are $3(2)\sigma$-level directional variations in $\Omega_bh^2$, $\Omega_ch^2$, $n_s$, $100\theta_\mathrm{MC}$, and $H_0$ $(\tau$ and $\ln(10^{10}A_s))$. Furthermore, the directional distributions of the parameters follow a dipole form to good approximation. 
The Bayes factor between the isotropic and anisotropic hypotheses is $0.0041$, strongly disfavouring the former.
The best-fit dipole axes for $\Omega_bh^2$, $\Omega_ch^2$, $n_s$, $100\theta_\mathrm{MC}$, and $A_s e^{-2\tau}$ all generally align with the mean direction of 
$\bm{V} \equiv (b = -5.6^{+17.0}_{-17.4}\degree, l = 48.8^{+14.3}_{-14.4}\degree)$,
which is roughly perpendicular to the dipole of the variation in fine structure constant, and is about $45\degree$ to the directions of the CMB kinematic dipole, CMB parity asymmetry, and polarization of QSOs.
Our results suggest either significant violation of the cosmological principle, or previously unknown systematic errors in the standard CMB analysis.
\end{abstract}
\maketitle

\section{Introduction}
The universe is assumed to be isotropic and homogeneous in large scales in the standard cosmological model, the $\Lambda$CDM model. 
However, there are observational data suggesting violation of the large-scale isotropy.
For example, previous studies indicate that the fine structure constant varies spatially at more than 4$\sigma$ confidence level using data from the Keck telescope and the Very Large Telescope \citep{PhysRevLett.107.191101,King_2012}.
There are significantly more left-handed than right-handed spiral galaxies 
\cite{SHAMIR201225}. 
There are directional variations in the expansion rate of the universe measured using Type IA Supernovae
 \cite{PhysRevD.64.063505,PhysRevD.86.083517}. 
There is a north-south asymmetry in the Cosmic Microwave Background (CMB) temperature anisotropy angular power $\mathcal{C}_\ell^{TT}$\cite{Eriksen_2007}\cite{PhysRevD.78.123520}.  
There are also anomalies in the low-$\ell$ part of ${\mathcal{C}_\ell}^{TT}$, such as the alignment of its quadrupolar and octupolar axes
\cite{Schild:2008fs}\cite{PhysRevD.69.063516}.  See \cite{preferred_review_2016} for a review of the preferred directions in cosmology.

In this paper, we look for spatial variations of the cosmological parameters from the CMB data. We perform Markov chain Monte Carlo (MCMC) fitting of cosmological parameters using the \textit{Planck} 2018 CMB data from half-skies centering on 48 different directions to look for their directional variations.

\section{Methodology}
There are six parameters \{$\Omega_b$, $\Omega_{c}$, $H_0$, $A_s$, $n_s$, $\tau$\} in the $\Lambda$CDM model, where $\Omega _{b, c}$ are the cosmological baryon and dark matter densities respectively, $H_0$ is the Hubble parameter at present, $A_s$ and $n_s$ are the primordial scalar amplitude and spectral index respectively, and $\tau$ is the reionization optical depth. 

\begin{figure}[hbt]
\includegraphics[width=\linewidth]{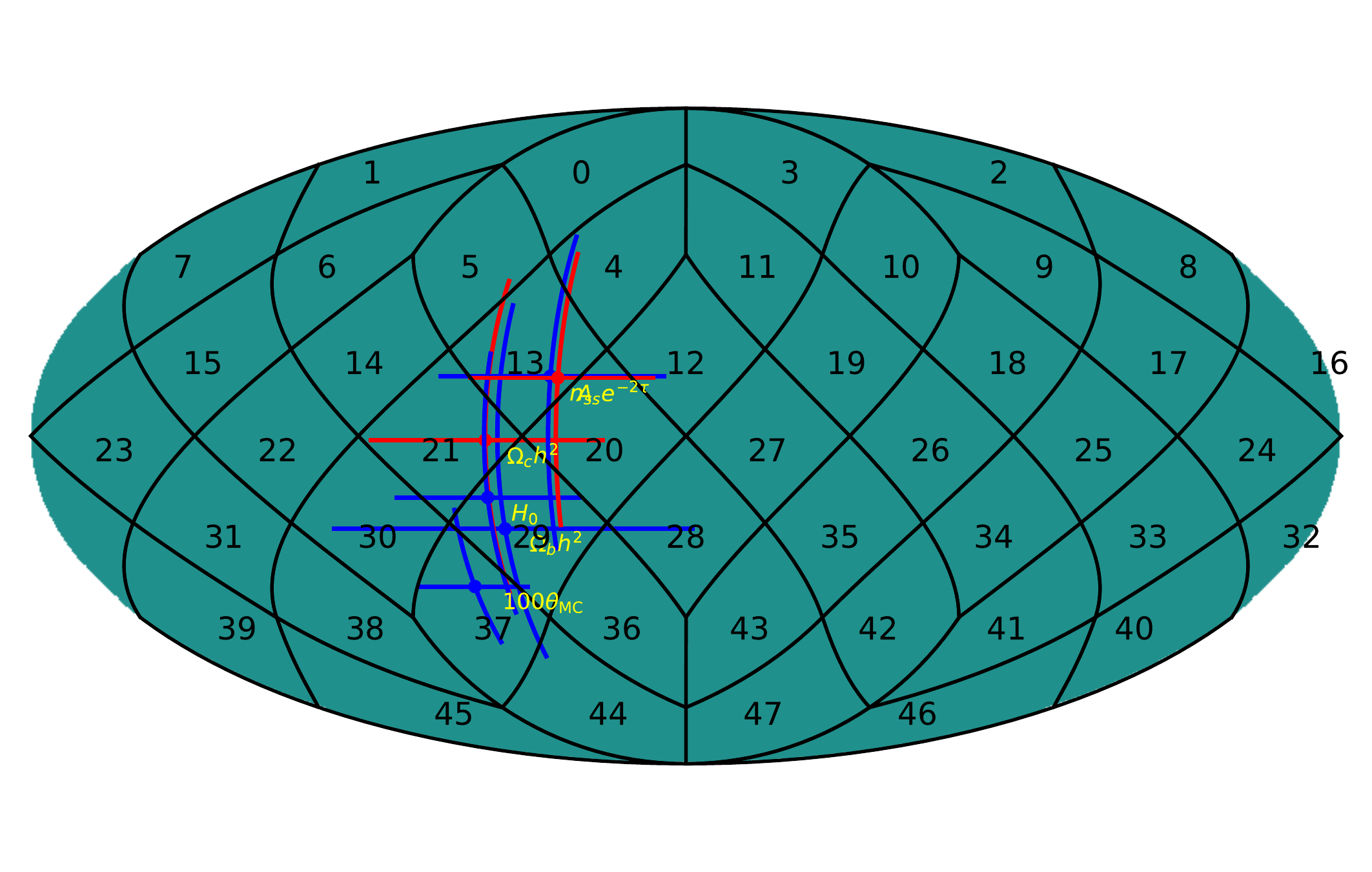}
\caption{Indices of the pixels used in the analysis. Fittings of CMB data are performed for 48 half-skies centering on these pixel centers.  We also put the best-fit dipole directions of the cosmological parameters on the figure. Red color indicates the dipole is flipped to the opposite direction.}
\label{fig:healpix-directions}
\end{figure}

We look for spatial variations of the cosmological parameters by applying an extra hemispherical mask on top of the original mask used in \textit{Planck}, to block off half of the sky in different directions, and compare the fitting results. We use the public code \texttt{CosmoMC} \footnote{http://cosmologist.info/cosmomc} to fit the CMB temperature angular power spectrum using the \textit{Planck} 2018 likelihood \cite{planck2018_likelihood}. Note that \texttt{CosmoMC} uses the parameter $100\theta_\mathrm{MC}$, which is approximately the ratio of the sound horizon to the angular diameter distance instead of $H_0$, since it is less correlated with other parameters. We follow this and use \{$\Omega_b$, $\Omega_{c}$, $100\theta_\mathrm{MC}$, $A_s$, $n_s$, $\tau$\} as the fitting parameters. We use the \textit{Planck} high-$\ell$ temperature and low-$\ell$ temperature and polarization data in the MCMC fittings. 
The data vectors and covariance matrices used in the likelihood function depend on the masks we use. Since we are additionally masking the opposite hemisphere of the direction under consideration, we have to recalculate the data vectors and covariance matrices.
We use \texttt{PolSpice} \footnote{http://www2.iap.fr/users/hivon/software/PolSpice/} to calculate the CMB anisotropy cross spectra using the appropriate sky maps, masks, and beam window functions. 
We then calculate the covariance matrices of different detector combinations by following the procedures in \cite{planck2015_likelihood}. In particular, we only calculate the covariances of the TT block (Eq.~C.2 of \cite{planck2015_likelihood}) since we only consider temperature data. The noise correlations are also considered by calculating the rescaling coefficients in Eq.~C.24 of \cite{planck2015_likelihood}. The covariance matrices also need to be corrected for the effects of the beam, pixel window function, and mask by Eq.~C.33 of \cite{planck2015_likelihood}. The excess variances induced by the point-source masks are also approximated by comparing empirical and theoretical power spectra variances (Appendix C.1.4 of \cite{planck2015_likelihood}). The final data vectors and covariances of different frequencies are the weighted average of the cross power spectra and the covariances of various detector combinations using the inverse of the diagonal elements of the covariance matrices as weights (Eq.~50 of \cite{planck2015_likelihood}).
We first make sure we can reproduce the standard \textit{Planck} MCMC results by following \cite{planck2018_likelihood}, using the original mask in the \textit{Planck} analysis, and following the steps mentioned above.
Then we apply the hemispherical masks according to the directions of the centers of the pixels in the HEALPix pixelization scheme, with the "RING" ordering. The parameter $N_{\mathrm{side}}$ is taken to be 2, and therefore there are 48 directions in total. The pixels are shown in Fig. \ref{fig:healpix-directions}.

\section{Results}

\begin{figure}[hbt]
\includegraphics[width=\linewidth,height=0.7\textheight,keepaspectratio]{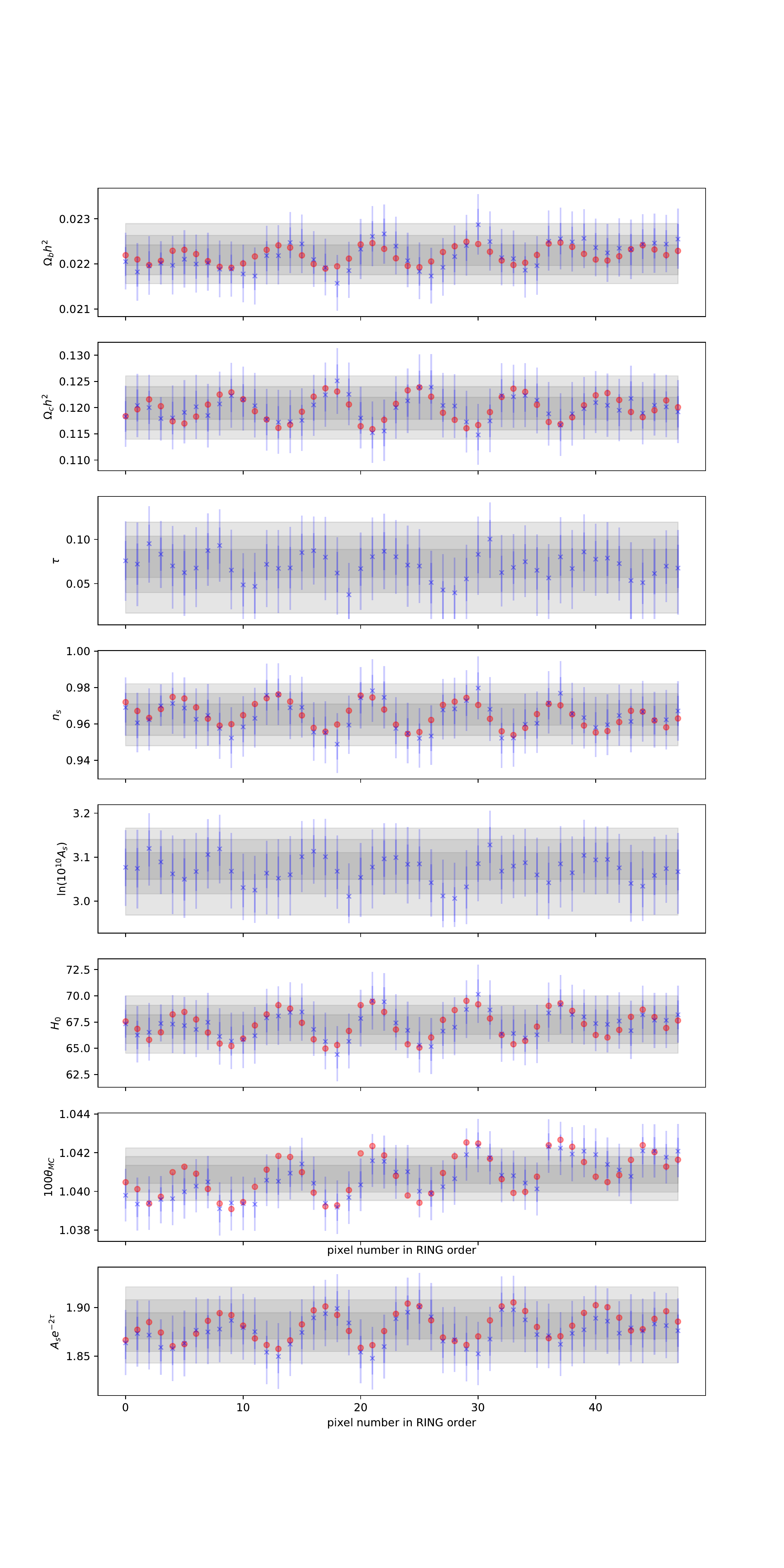}
\caption{Mean (denoted by crosses),  68\% CI (denoted by dark blue bars), and 95\% CI (denoted by light blue bars) of the cosmological parameters from the MCMC fittings using \textit{Planck} CMB temperature data and assuming $\Lambda$CDM model. The results are obtained with hemispherical masks applied in addition to the original mask according to the directions in Fig.~\ref{fig:healpix-directions}. The grey bands are the 68\%, 95\%, and 99.7\% CI from the full-sky case. The red dots are calculated using the means of the dipoles from Table \ref{table:dipole_nside2}.}
\label{fig:scatter_all_par}
\end{figure}

\begin{figure}[hbt]
\includegraphics[width=\linewidth]{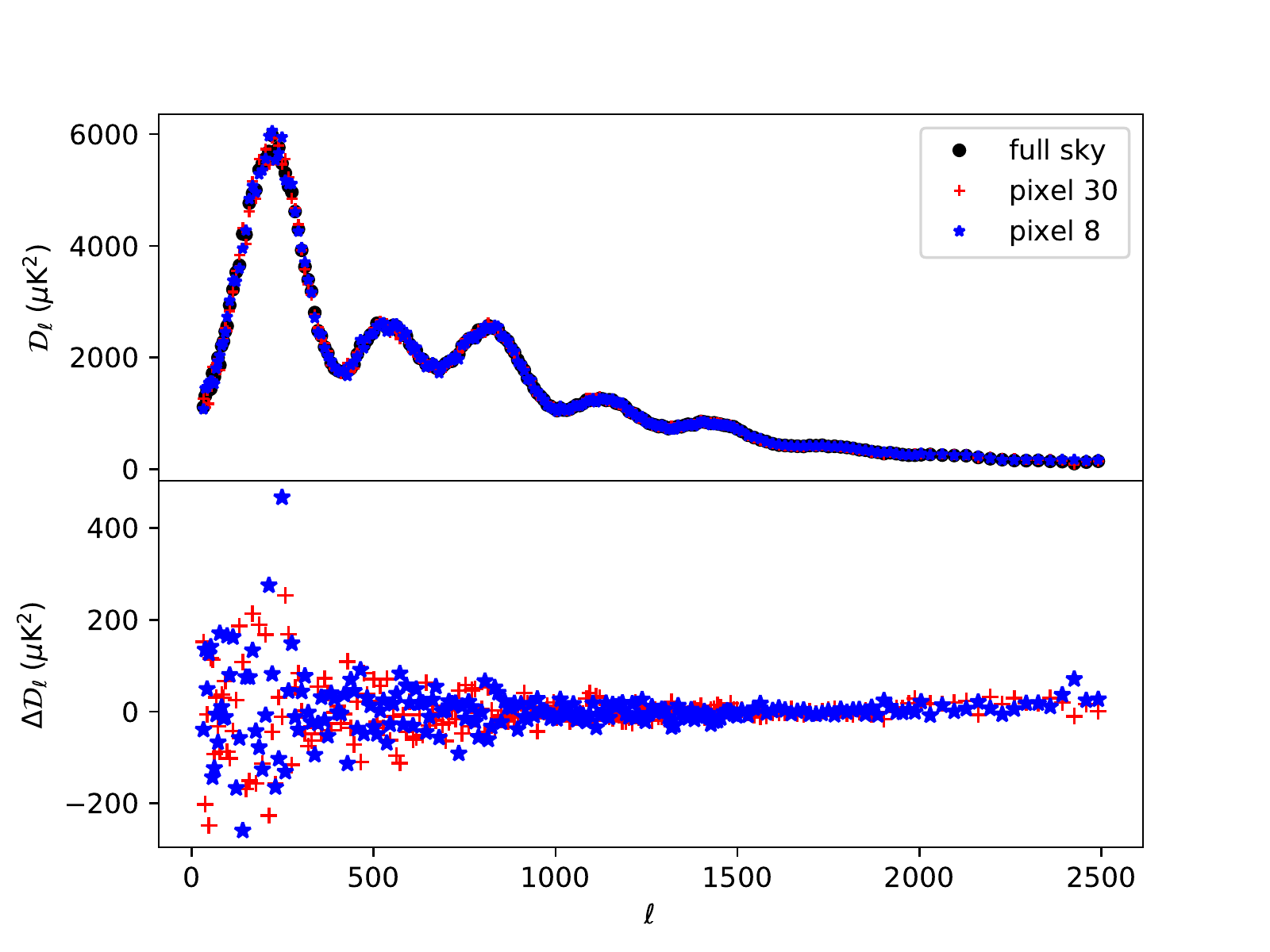}
\caption{Top: Binned CMB anisotropy power spectra $\mathcal{D}_\ell\equiv \frac{\ell(\ell+1)}{2\pi}\mathcal{C}_\ell$ of the full-sky and half-skies centering at pixels 8 and 30. Bottom: Differences between the power spectra of the half-skies and the full-sky.}
\label{fig:cl_full_30_8}
\end{figure}

The $\Lambda$CDM fitting results of the \textit{Planck} data are shown in
Fig.~\ref{fig:scatter_all_par}. 
The half-sky at pixel 8 (30) has the smallest (largest) value of $\theta_\mathrm{MC}$ from the MCMC results. To demonstrate the effect of the additional masks, we plot the binned $\mathcal{D}_\ell$ of the full-sky, half-sky at pixel 8, and half-sky at pixel 30 in Fig.~\ref{fig:cl_full_30_8}. 

There are directional variations of $\Omega _{b,c}h^2$, $n_s$, $100\theta_\mathrm{MC}$, and $A_se^{-2\tau}$ up to 3 $\sigma$. For example, the best-fit values of $n_s$ change from $0.9488 \pm 0.008$ (pixel 18) to $0.9797 \pm 0.0087$ (pixel 30), whereas the full-sky value is $0.9655 \pm 0.0062$.
The mean value of $100\theta_\mathrm{MC}$ at the direction of pixel 8 is 1.0391, deviating by more than 3$\sigma$ away from the full-sky value $1.040878\pm 0.00047$.
In some of the directions, the lower bounds of the 68\% CI of $\tau$ reach the prior limit due to the variation of their means.

\begin{figure}[hbt]
    \centering
    \includegraphics[width=\columnwidth]{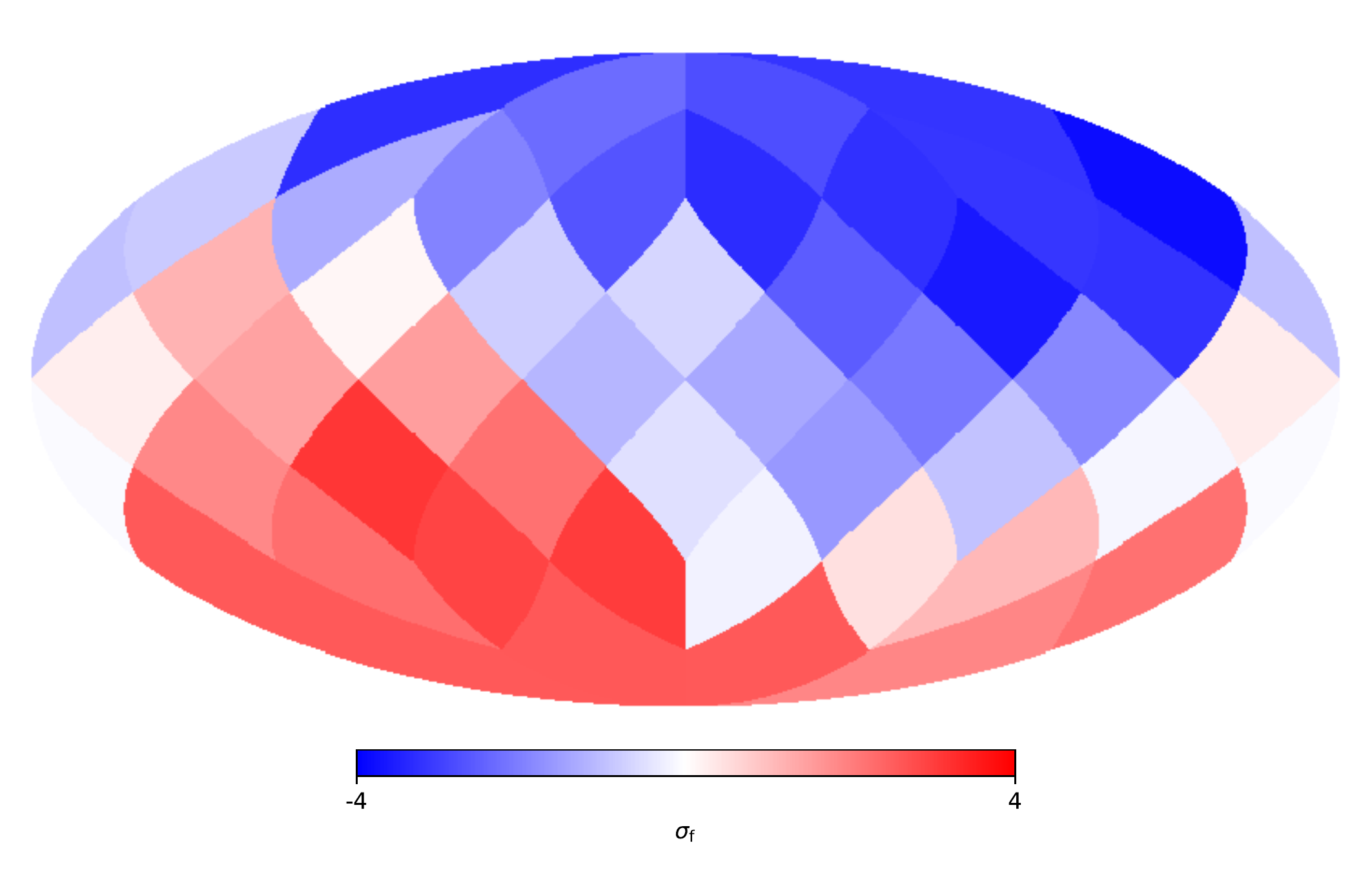}
    \caption{Deviations of the means of $\theta_\mathrm{MC}$ with respect to the full-sky value in terms of the full-sky standard deviation $\sigma_\mathrm{f}$, fitted using data on the hemisphere centered at different directions.}
    \label{fig:theta_sky_nside2}
\end{figure}

Following \cite{PhysRevD.86.083517}, 
we fit the directional variations of the cosmological parameters with respect to the full-sky means in Fig.~\ref{fig:scatter_all_par} to a dipole form of $\bm{d}\cdot \hat{n}$, where $\bm{d}$ is the dipole vector and $\hat{n}$ is the unit vector pointing to the 48 directions. By Bayes' theorem, 
\begin{align}
     p(\bm{d}|\{D_\mathrm{h}\})&\propto p(\{D_\mathrm{h}\}|\bm{d})p(\bm{d}),
\end{align}
where $\{D_\mathrm{h}\}$ is the hemi-sky data at different directions.
We assume the parameters are normally distributed for easier computation. Since the distributions of $\tau$ and $\ln A_s$ are not gaussian in some of the directions, we consider the combination $A_s e^{-2\tau}$ instead. The detailed fitting procedure is presented in the Appendix. The results are presented in Table \ref{table:dipole_nside2}. 
The directions of the dipoles are also shown in Fig. \ref{fig:healpix-directions}, and the deviations of the means of $100\theta_\mathrm{MC}$ at different directions with respect to the full-sky value are shown in Fig. \ref{fig:theta_sky_nside2}. From Table \ref{table:dipole_nside2}, we can see that some of the dipoles are significant. For example, $d_y$ of $100\theta_\mathrm{MC}$ is about 4$\sigma$ away from zero. 

We can test the hypothesis $\Hi$ that the universe is isotropic by using the hemispherical data. Let $\Ha$ be the alternative hypothesis that the universe is not isotropic.
By Bayes' theorem,
\begin{align}
     p(\Hi|\{D_\mathrm{h}\})&=\frac{p(\{D_\mathrm{h}\}|\Hi)p(\Hi)}{p(\{D_\mathrm{h}\})}.
\end{align}
The Bayes factor $K$ is the ratio between the probabilities of the two hypotheses:
\begin{align}
    K\equiv \frac{p(\Hi|\{D_\mathrm{h}\})}{p(\Ha|\{D_\mathrm{h}\})}&=\frac{p(\{D_\mathrm{h}\}|\Hi)p(\Hi)}{p(\{D_\mathrm{h}\}|\Ha)p(\Ha)}.
\end{align}
The detailed calculations can be found in the Appendix. We get $K=0.0041$, with 95\% of samples being smaller than 0.046 from bootstrapping. Therefore, we conclude that the CMB data provides strong evidence for $\Ha$.

\begin{figure}
    \centering
    \includegraphics[width=\linewidth]{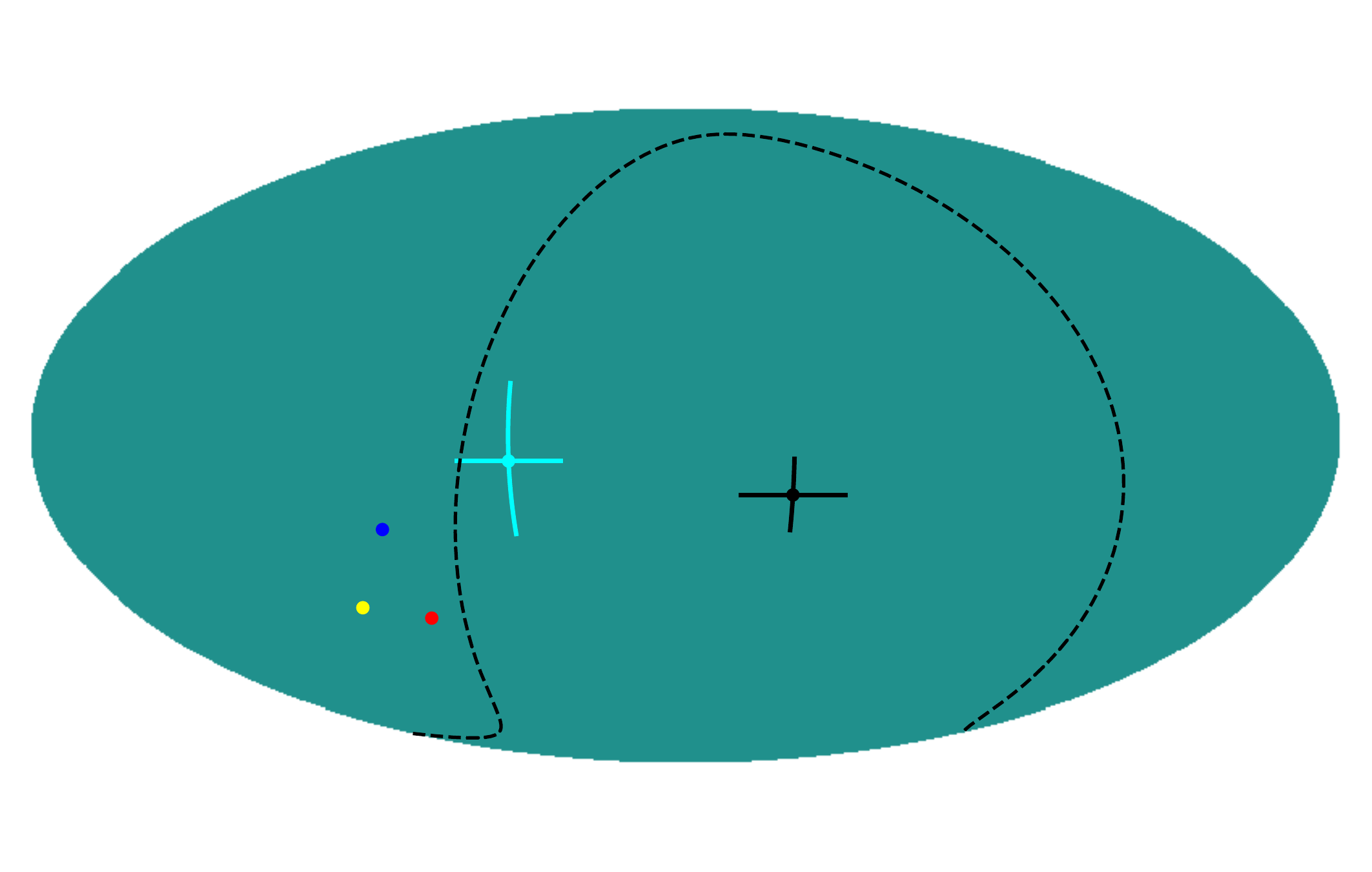}
    \caption{Directions of $\bm{V}$ (cyan), the dipole of the fine structure constant \cite{King_2012} (black), the CMB kinematic dipole (red), the quasar polarization vector (blue), and the CMB parity asymmetry (yellow). The latter three are flipped to their opposite directions, and are taken from \cite{preferred_review_2016}. The dashed line shows the ring $90\degree$ away from the dipole of the fine structure constant.}
    \label{fig:directions}
\end{figure}

Interestingly, the best-fit dipole axes for $\Omega_ch^2$, $\Omega_bh^2$, $100\theta_\mathrm{MC}$, $n_s$, and $A_s e^{-2\tau}$ all align around Pixels 21 and 30, with a mean direction of $\bm{V} \equiv (b = -5.6^{+17.0}_{-17.4}\degree, l = 48.8^{+14.3}_{-14.4}\degree)$, which is roughly perpendicular (at 77$\degree$) to the dipole of the fine structure constant \citep{King_2012}. $\bm{V}$ is roughly $45\degree$ away from the directions of the CMB kinematic dipole, CMB parity asymmetry, and polarization of QSOs \citep{preferred_review_2016}. These directions are plotted in Fig.~\ref{fig:directions}, which shows that $\bm{V}$ does not align with other directions.
Notice that we do not include $H_0$ here because it is highly correlated with $100\theta_\mathrm{MC}$. 

\begin{table*}[hbt]
    \centering
    \begin{ruledtabular}
        \begin{tabular}{llll}
            parameter & $d_x$ & $d_y$ & $d_z$\\
            \hline
$\Omega_b h^2$ & $0.00017\pm 0.00023$ & $0.00022\pm 0.00025$ & $-0.00012\pm 0.00032$\\
$\Omega_c h^2$ & $-0.0023^{+0.0021}_{-0.0022}$ & $-0.0034\pm 0.0022$ & $0.0001\pm 0.0031$\\
$n_s$ & $0.0087\pm 0.0055$ & $0.0067^{+0.0058}_{-0.0057}$ & $0.0028^{+0.0088}_{-0.0087}$\\
$H_0$ & $1.26\pm 0.96$ & $1.84^{+0.96}_{-0.94}$ & $-0.6\pm 1.4$\\
$100\theta_{MC}$ & $0.00063\pm 0.00045$ & $0.00133^{+0.00035}_{-0.00034}$ & $-0.00105\pm 0.00058$ \\
$A_s e^{-2\tau}$ & $-0.0189\pm 0.0092$ & $-0.014\pm 0.011$ & $-0.006\pm 0.015$\\
        \end{tabular}
    \end{ruledtabular}
    \caption{The mean values and 68\% CI from the fitting results of the directional variations of cosmological parameters with respect to the full-sky mean in Fig.~\ref{fig:scatter_all_par} to the dipole form $\bm{d}\cdot \hat{n}$. }
    \label{table:dipole_nside2}
\end{table*}

\begin{figure}[hbt]
    \centering
     \includegraphics[width=\linewidth]{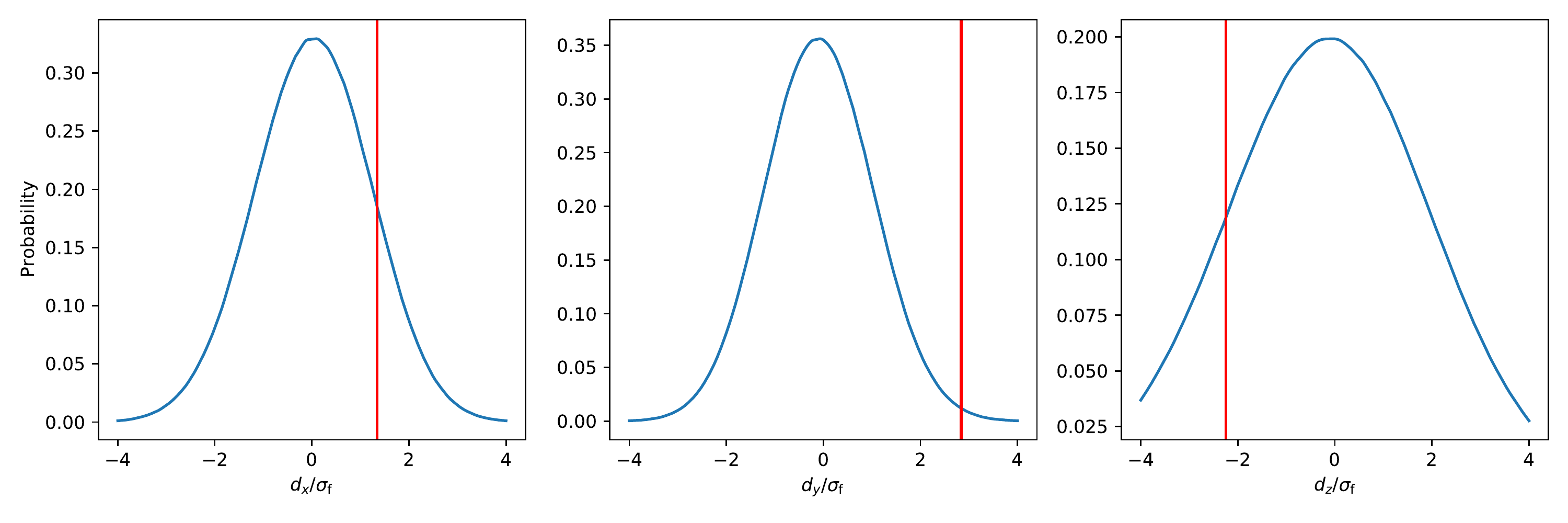}
    \caption{Smoothed probability distributions of the dipole components ($x$, $y$, and $z$ in the left, middle and right panels, respectively) of the parameter $100\theta_\mathrm{MC}$ from 100 simulated sky maps, using $N_\mathrm{side} = 1$, or 12 directions. The red vertical lines represent the mean values from the real sky. The distributions are smoothed by kernel density estimation using gaussian kernels and bandwidths estimated by Scott's Rule.}
    \label{fig:mc_dipole}
\end{figure}

To test our procedure, we have also performed the above analysis using the \textit{Planck} \texttt{FFP8} simulated sky maps, including 100 different sets of CMB signal and noise maps, but for $N_\mathrm{side} = 1$, or 12 directions only.
The simulated CMB sky maps from Planck contain the lensing, Rayleigh scattering, and Doppler boosting effects convolved with the beams. The simulated noise sky maps include time variations in the noise power spectral density of each detector \cite{planckffp8}.
The simulated sky maps are masked in the same way as the real sky maps. The smoothed distributions of the mean values of the fitted dipole components of the parameter $100\theta_\mathrm{MC}$ are shown in Fig.~\ref{fig:mc_dipole}, which are consistent with having no dipole. This shows that our analysis procedure does not bias the estimation of the parameters. We also added the kinematic dipole term to the simulated skies, and this does not change the results.

To quantify the alignment of the dipoles of the parameters, we consider a modified version of the spherical variance of the directions of the best-fit dipoles, defined as
\begin{align}
    S  = 1-\frac{1}{N} \sum_{i\neq j} \frac{(\bm{d}_i\cdot\bm{d}_j)^2}{\left\lVert \bm{d}_i \right\rVert^2 \left\lVert \bm{d}_j \right\rVert^2},
\end{align}
where the sum is over all $N$ combinations of the dipole directions. $S$ measures how good the dipole directions align with each other while ignoring their signs, so that two vectors pointing to opposite directions are still considered perfectly aligned.
Let the hypothesis that the dipoles align better (worse) than the standard $\Lambda$CDM model be $\mathcal{H}_1$ ($\mathcal{H}_0$). We calculate the Bayes factor between $\mathcal{H}_1$ and $\mathcal{H}_0$. We use the distributions of the dipoles of the 100 \texttt{FFP8} simulated sky maps to estimate the distribution of $S$ in an isotropic universe, $p_\mathrm{i}(S)$. The dipoles are considered to have a better alignment if the corresponding $S$ is smaller than the median of $p_\mathrm{i}(S)$, $S_0$. Consider $p(\mathcal{H}_1|\{D_\mathrm{h}\})$:
\begin{align}
    p(\mathcal{H}_1|\{D_\mathrm{h}\}) &\propto p(\{D_\mathrm{h}\}|\mathcal{H}_1)p(\mathcal{H}_1)\\
    &=\int \dd \bm{S}\, p(\{D_\mathrm{h}\}|S)p(S|\mathcal{H}_1)p(\mathcal{H}_1)\\
    &=\int_{0}^{S_0} \dd \bm{S}\, p_D(S)p(\mathcal{H}_1),
\end{align}
where $p_D(S)$ is the distribution of $S$ calculated from the posteriors of the dipoles. $p(\mathcal{H}_0|\{D_\mathrm{h}\})$ can be calculated similarly. The Bayes factor is 1.67, which is ``barely worth mentioning".

The Hubble parameter $H_0$ in units of $\unit{km\,s^{-1}\,Mpc^{-1}}$ have means of 64.4 and 70.1 and standard deviations of 1.3 and 1.4 in the two extreme directions of Pixels 18 and 30, respectively. The directional dependence of $H_0$ has a comparable magnitude as the difference between the CMB and local measurements of $H_0$  \cite{Riess_2016,Riess_2018}, and our results may have implications on this famous tension. Our results may suggest significant deviations from the $\Lambda$CDM model, or previously unknown systematic errors in the standard CMB analysis to extract cosmological parameters. The masking procedure and the CMB kinematic dipole cannot introduce such systematic errors since the analysis results using simulated skies are consistent with no dipole.

\section{Conclusions}
We performed statistical analysis of the angular distribution of the cosmological parameters by adding hemispherical masks at different directions to the CMB data. The directions were chosen to be the center of the pixels in the \texttt{HEALPix} pixelization scheme with $N_\mathrm{side}=2$. We used the \textit{Planck} 2018 high-$\ell$ temperature and low-$\ell$ temperature and polarization data together with \texttt{CosmoMC} to get the posteriors of the cosmological parameters by MCMC.

There are $3(2)\sigma$-level directional variations in $\Omega_bh^2$, $\Omega_ch^2$, $n_s$, $100\theta_\mathrm{MC}$, and $H_0$ $(\tau$ and $\ln(10^{10}A_s))$. Furthermore, the cosmological parameters follow to good approximation a dipole form, with $100\theta_\mathrm{MC}$ being the most significant. The dipole axes for $\Omega_ch^2$, $\Omega_bh^2$, $100\theta_\mathrm{MC}$, $n_s$, and $A_s e^{-2\tau}$ all align around Pixels 21 and 29, which is about $45\degree$ away from the directions of the CMB kinematic dipole, CMB parity asymmetry, and polarization of QSOs, and is roughly perpendicular to the dipole of the variation of the fine structure constant.
By considering $100\theta_\mathrm{MC}$ only, we calculated the Bayes factor $K$ between the isotropic hypothesis in which $100\theta_\mathrm{MC}$ is a constant over the whole sky and the alternate hypothesis in which $100\theta_\mathrm{MC}$ has a dipole distribution. We found that $K\approx 0.0041$, which means that the isotropic hypothesis is strongly disfavored. We also performed the analysis using 100 simulated sky maps from \textit{Planck} \texttt{FFP8}, which include the lensing, Rayleigh scattering, Doppler boosting effects, and are convolved with the beams. These effects do not give rise to any significant dipole in the cosmological parameters. Our results also indicate that the masks used in the Planck CMB analysis and the CMB kinematic dipole are not the causes of the significant dipoles in the cosmological parameters we have found. This suggests that there are significant violations of the cosmological principle, or previously unknown systematic errors in the standard CMB analysis that are not considered in the \texttt{FFP} simulations.

\acknowledgments
This work is partially supported by the Research Grant Council of the Hong Kong Special Administrative Region, China (Project No.s 14301214, AoE/P-404/18). We acknowledge the support of the CUHK Central High Performance Computing Cluster, on which the computation in this work has been performed. We thank K.P. Chan for helpful discussion of the idea.

\appendix*
\section{Fittings of the dipole and calculations of the Bayes factor $K$}
\label{appendix}

To fit the dipole, we need to specify the form of the likelihood and the prior.
From the full-sky MCMC, the marginalized mean and standard deviation of the parameter $p$ are $\bar{p}$ and $\sigma_{p,\mathrm{f}}$, respectively. The posterior of the full-sky MCMC is approximately gaussian: $p(p|D_\mathrm{fs})\approx g(p|\bar{p}, \sigma_{p,\mathrm{f}})$. We approximate the covariance matrix $\Sigma$ between the half-sky data by the MCMC results of 100 simulated skies in 12 directions of $N_\mathrm{side}=1$ according to the method in \citep{cov_struct}. In this method, we place the stationary processes in the north and south pole. Both the stationary covariance function and the corresponding weights are of the form of squared exponential. The smoothing parameter is 0.5.
The joint posterior distribution of $p$ at different directions is then given by  $p(\bm{p}|\{D_\mathrm{h}\})\approx g(\bm{p}|{\bm{\bar{p}_\mathrm{h}}}, \bm{\Sigma})$.
We assume that the mean values of $p$ at different directions follow a dipole distribution $p_d(\hat{n})=\bar{p}+\hat{n}\cdot(d_x\hat{x}+d_y\hat{y}+d_z\hat{z})$. Assuming the same covariance $\Sigma$ between different directions, $\bm{p}$ should follow the distribution $p(\bm{p}|\bm{d})\approx g(\bm{p}|\bm{p_d}, \bm{\Sigma})$. 
The likelihood term is:
\begin{align}
    p(\{D_\mathrm{h}\}|\bm{d})
    &=\int \dd \bm{p}\, p(\{D_\mathrm{h}\}|\bm{p})p(\bm{p}|\bm{d})\\
    &\propto \int \dd \bm{p}\, p(\bm{p}|\{D_\mathrm{h}\})p(\bm{p}|\bm{d})\\
    &\approx \int \dd \bm{p}\, g(\bm{p}|{\bm{\bar{p}_\mathrm{h}}}, \bm{\Sigma})g(\bm{p}|\bm{p_d}, \bm{\Sigma})\\
    \label{eq:prod_int_2gaussian}
    &\propto \exp\left(-\frac{1}{4}(\bm{\bar{p}_\mathrm{h}}-\bm{p_d})^T\bm{\Sigma}^{-1}(\bm{\bar{p}_\mathrm{h}}-\bm{p_d})\right).
\end{align}
The priors of the dipole components $d_{x,y,z}$ are uniform between $\pm 4\sigma_\mathrm{f}$. Hence the posteriors of the dipole components are truncated three dimensional normal distribution.

We focus on the parameter $\theta_\mathrm{MC}$ for the hypothesis testing. To compute $K$, we need to calculate $p(\{D_\mathrm{h}\}|\Hi)$ and $p(\{D_\mathrm{h}\}|\Ha)$. In the previous fittings of the dipoles, we have implicitly assumed the anisotropic hypothesis $\Ha$ where the parameter follows a dipole form. $p(\{D_\mathrm{h}\}|\Ha)$ is given by the marginalization over $\bm{d}$:
\begin{align}
    p(\{D_\mathrm{h}\}|\Ha) &= \int \dd \bm{d}\,  p(\{D_\mathrm{h}\},\bm{d}|\Ha)\\
    &= \int \dd \bm{d}\,  p(\{D_\mathrm{h}\}|\bm{d})p(\bm{d}).
\end{align}

\begin{figure}[ht]
    \centering
    \includegraphics[width=\linewidth]{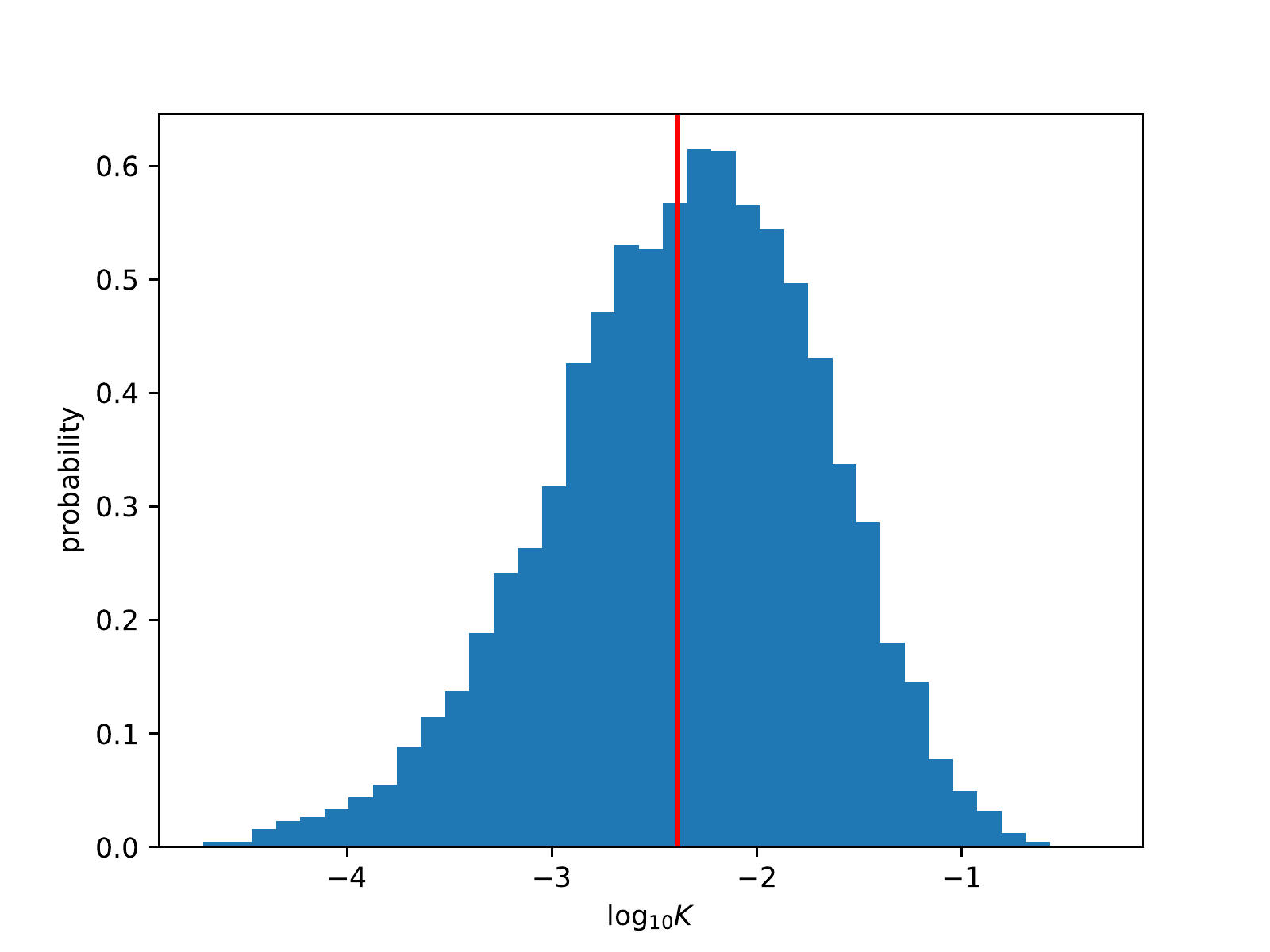}
    \caption{Histogram of $K$ from bootstrapping. The mean value is 0.0041, and 95\% of the samples are smaller than 0.046. The red line indicates the value of $K$ calculated from the empirical $\bm{\Sigma}$.}
    \label{fig:bootstrap_K}
\end{figure}

The calculation of $p(\{D_\mathrm{h}\}|\Hi)$ is similar. Instead of $\bm{d}$, we consider a constant $A$ over the whole sky: $p(\bm{p}|A)\approx g(\bm{p}|\bm{\bar{p}+A}, \bm{\Sigma})$. The prior of $A$ is uniform between $\pm 4\sigma_\mathrm{f}$.
We choose the same prior for $\Hi$ and $\Ha$: $p(\Hi)=p(\Ha)=0.5$. Since $\bm{\Sigma}$ may not be accurate, we use bootstrapping to estimate the reliability of the calculation. Bootstrapping shows that $K$ is not biased, and 95\% of the samples are smaller than 0.046. The histrogram from bootstrapping is shown in Fig.~\ref{fig:bootstrap_K}.

\clearpage
\bibliographystyle{apsrev4-1}
\bibliography{long_abbreviation,references}

\end{document}